\documentclass[a4paper,fleqn,usenatbib]{mnras}

\usepackage{newtxtext,newtxmath}

\usepackage[T1]{fontenc}
\usepackage{ae,aecompl}
\usepackage{pdflscape}
\usepackage{subfig}

\usepackage{graphicx}	
\usepackage{amsmath}	
\usepackage{amssymb}	
\usepackage{color}

\title[Performance analysis of periodogram tools]{Searching for planetary signals in Doppler time series: a performance evaluation of tools for periodograms analysis}

\author[M. Pinamonti et al.]{
Matteo Pinamonti,$^{1,2}$\thanks{E-mail: matteo.pinamonti@phd.units.it}
Alessandro Sozzetti,$^{3}$
Aldo S. Bonomo$^{3}$
and Mario Damasso$^{3}$
\\
$^{1}$Dipartimento di Fisica, Universit\`a degli Studi di Trieste, via G. B.Tiepolo 11, I-34143 Trieste, Italy\\
$^{2}$INAF - Osservatorio Astronomico di Trieste, via G. B. Tiepolo 11, I-34143, Trieste, Italy\\
$^{3}$INAF - Osservatorio Astrofisico di Torino, Via Osservatorio 20, I-10025 Pino Torinese, Italy
}

\date{Accepted XXX. Received YYY; in original form ZZZ}

\pubyear{2017}

\begin{document}
\label{firstpage}
\pagerange{\pageref{firstpage}--\pageref{lastpage}}
\maketitle

\begin{abstract}
We carry out a comparative analysis of the performance of three algorithms widely used to identify significant periodicities in radial-velocity (RV) datasets: 
the Generalised Lomb-Scargle Periodogram (GLS), its modified version based on Bayesian statistics (BGLS), and the multi-frequency periodogram scheme called FREquency 
DEComposer (FREDEC). We apply the algorithms to a suite of numerical simulations of (single and multiple) low-amplitude Keplerian RV signals induced by low-mass 
companions around M-dwarf primaries. 
The global performance of the three period search approaches is quite similar in the limit of an idealized, best-case scenario (single planets, circular orbits, 
white noise). However, GLS, BGLS and FREDEC are not equivalent when it comes to the correct identification of more complex signals (including correlated noise of stellar origin, eccentric orbits, multiple planets), with variable degrees of efficiency loss as a function of system parameters and degradation in completeness and reliability levels.
The largest discrepancy is recorded in the number of false detections: the standard approach of residual analyses adopted for GLS and BGLS translates in large fractions of false alarms ($\sim30\%$) in the case of multiple systems, as opposed to $\sim10\%$ for the FREDEC approach of simultaneous multi-frequency search. 
Our results reinforce the need for the strengthening and further development of the most aggressive and effective {\it ab initio} strategies for the 
robust identification of low-amplitude planetary signals in RV datasets, particularly now that RV surveys are beginning to achieve sensitivity to potentially habitable 
Earth-mass planets around late-type stars. 
\end{abstract}




\section{Introduction}

The growing evidence from transit (e.g., Kepler) and radial-velocity (e.g, HARPS, HARPS-N) surveys points towards a high occurrence rate of 
low-mass ($\leq30$  M$_\oplus$), small-size ($\leq3$ R$_\oplus$) planets \citep[e.g.,][]{mayor2011,howard2013}, with a large fraction of late-type M dwarfs 
hosting habitable-zone terrestrial-type companions (see, e.g., \citealt{winnfab15}, and references therein). The combined statistical inferences from HARPS and Kepler indicate that 
planets in the range between Super Earths and Neptunes are not only very common, but they are often found in multiple systems, tightly packed close to the central star, 
and almost perfectly coplanar when seen in transit \citep[e.g.,][]{batalhaetal2013,roweetal2014,fabrycky2014}. The observational evidence is posing a formidable 
challenge for planet formation and evolution models, but it is also inducing a fundamental change of perspective in radial velocity (RV) observing strategy. 
The ubiquitousness of multiple systems with low-mass components requires a very significant investment of observing time for a proper modeling of the complex signals. 
Usually, multi-year campaigns with hundreds of RVs are presented in discovery announcements of Super Earths and Neptune-like planets \citep[e.g.,][]{bonfils2013,astudef2015}.
In addition, the analysis of low-amplitude signals is often complicated by stellar activity, that can induce false positive signals mimicking the RV signature of 
a low-mass planet, and induce systematic effects comparable in magnitude to (and even exceeding) the amplitudes of the sought after Keplerian signals \citep[e.g.,][]{pepeetal2013}. 

In the search for low-mass planets with spectroscopic surveys, the first step in the investigation of unevenly spaced RV time series relies on the identification 
of statistically significant periodic signals via a variety of implementations of a periodogram analysis. The Lomb-Scargle periodogram \citep[LS,][]{lomb1976,scargle1982}, 
which performs a full sine-wave fit over a large grid of trial frequencies, has historically been the first tool adopted for the task. More recently some authors have 
extended the LS formalism to include weights for the measurement errors and constant offsets for the data in the Generalized Lomb-Scargle (GLS) periodogram \citep{zechkur2009}, 
and generalizations based on Bayesian probability theory in the Bayesian Lomb-Scargle and Bayesian Generalized Lomb-Scargle periodograms (BLS, BGLS) \citep{bretthorst2001,mortier2015}. 
Due to the high fraction of low-mass multiple-planet systems, and also to the presence of activity related signals in the data, the correct identification of multiple, 
low-amplitude signals is of course a central issue in RV time series analysis as applied to exoplanet science. However, all the above algorithms fit only a single sine-wave, 
or keplerian signal, and multiple signals must be detected via subsequent fits and residual analysis. To overcome some of the shortcomings of standard periodograms 
when dealing with data containing two or more periodicities \citet{baluev2013fre} has developed the multi-frequency periodogram FREquency DEComposer (FREDEC).

In this work we expand on the study by \citet{mortier2015} and carry out a set of detailed numerical experiments aimed at 1) gauging the relative effectiveness of the 
GLS, BGLS, and FREDEC algorithms, including completeness and false positives, and 
2) understand their biases and limitations when applied to the systematic search of single and multiple low-amplitude periodic 
signals produced by low-mass companions, using M dwarfs as choice of reference for the central star. The performance evaluation in the presence of representative 
complex signals element constitutes a novel analysis that has not been undertaken before, to our knowledge. This comparative study should not be interpreted 
as a way of ranking the intrinsic effectiveness of a periodogram analysis method against another. Rather, it has to be seen as one of the steps 
that will help towards the definition and implementation of the most aggressive and effective strategies (e.g., \citealt{dumusque2017,hara16}, and references therein) 
for a robust identification of terrestrial planetary systems with state-of-the-art instrumentation (e.g. HARPS, HARPS-N) that guarantees meter-per-second accuracy, 
as well as next-generation facilities for extreme precision RV measurements, such as ESPRESSO. In section \ref{sim.set} we describe the numerical setup adopted in our study, 
while the main results of our suite of simulations are presented in Section \ref{sing.res}. We provide a summary and discussion of our findings in Section \ref{discuss}.

\section{Simulation setup}
\label{sim.set}

\subsection{Assumptions and Caveats}

The suite of simulated catalogs of RV observations described below and utilized in the analysis has been produced using a set of working assumptions and simplifications. 
In particular: 

\begin{itemize}

\item The comparative performance avaluation of GLS, BGLS, and FREDEC is expressed in terms of the dependence of the efficiency of signal recovery (parameterized through the theoretical 
false alarm probability FAP) on the main orbital elements it is expected to depend upon, i.e. orbital period $P$, eccentricity $e$, RV semi-amplitude $K$, and the 
'signal-to-noise' ratio $K/\sigma$, 
where $\sigma$ is the single-measurement RV error. The adoption of the theoretical FAP rather than its calculation via bootstrap methods was dictated by the need to keep processing 
time within reasonable boundaries given the computational resources at our disposal;  

\item RV measurements are affected by a random (Gaussian) noise component. In one experiment, a simple synthetic stellar activity signal was added to the RV data. 
This was done as a metric of comparison with recent literature works, while a full-scale study of the effect of correlated stellar noise is left for future developments. 
We also did not consider the presence of outer companions, stellar or planetary, that would introduce long-term RV drifts; 

\item Up to two low-mass planets where simulated. The growing evidence for the existence of compact multiple systems with a number of planets significantly exceeding 2 
naturally calls for relaxation of this assumption. Our aim is to identify proxies for interpreting in a simple manner any differences in behaviour of the three algorithms 
that might arise in the case of two-planet systems that might be used in a future work for easing the understanding of the efficiency of periodogram analyses carried out with 
a variety of methods in cases of even more complex RV signals. 

\item In the simulations we included the elements of the window function appropriate for reproducing the gaps in the data due to the seasonality of the observations 
as well as the alternation between day and night. The number of RV measurements per season (a few tens) was that typical of current RV surveys, rather than that used 
in very intensive observational campaigns (with hundreds of datapoints) focused on few targets. No prescriptions were made for either the generation of gaps in the data due to long 
stretches of bad weather, or the generation of RVs with large uncertainties as if obtained under not optimal weather conditions.

\end{itemize}

\subsection{Synthetic catalogs}
\label{synt.cat}

We created several catalogs of synthetic RV time series. 
Each time series consists of $N$ radial velocity measurments $y_i$ distributed over a number $N_s$ of observing seasons, 
their respective times $t_i$, and the associated errors $\sigma_i$ ($i=1,\dots,N$). The 
Keplerian RV signal induced by the $j^{th}$ planetary companion is evaluated through the standard formula:

\begin{equation}
y_j(t)= K_j [e_j \cos \omega_j + \cos (\nu_j (t) + \omega_j)] + \gamma,
\label{eq.rv}
\end{equation}
with $\omega_j$ the longitude of periastron, $\nu_j(t)$ the true anomaly, and $\gamma$ a constant offset. One obtains $\nu_j(t)$ in terms of $e_j$ 
and the eccentric anomaly $E_j(t)$ as:
\begin{equation}
\tan {\nu_j(t) \over 2} = \sqrt{{1+ e_j \over 1- e_j}} \tan{E_j(t) \over 2},
\end{equation}
with $E_j(t)$ determined via iterative solution of Kepler's equation:
\begin{equation}
E_j(t) - e_j \sin E_j(t) = M_j(t) = 2 \pi {t - T_{0,j} \over P_j},
\label{eq.kep}
\end{equation}
where $M_j(t)$ is the mean anomaly and T$_{0,j}$ the time of periastron passage. From the orbital parameters we can recover the planets' 
minimum mass $M_{p,j} \sin i_j$ using the relation:
\begin{equation}
\label{eq.km}
M_{p,j} \sin i_j \propto K_j  P_j ^{1 / 3} M_\star ^{-2/3} (1 - e_j^2)^{1/2},
\end{equation}
where $M_\star$ is the mass of the primary. The value of $M_\star$ and $\gamma$ were kept constant to $M_{\star} = 0.5 M_{\sun}$ and $\gamma= 0.0$ m s$^{-1}$, respectively, 
throughout our study.

The instrumental noise was modeled as purely white, with the single-measurement error $\sigma_i$ drawn from a Gaussian distribution with standard deviation 
of 1.5 m s$^{-1}$, which is representative of typical values of internal errors in Doppler time-series of relatively bright M dwarfs. 
The generation of the synthetic systems and relative RV signals was carried out with a set of prescriptions detailed below. 

\subsubsection{Single-planet circular orbits catalog}
\label{c.cat}

The first catalog consists of 10\,000 synthetic systems composed of a single companion on a circular orbit ($e=0.0$). 
The orbital parameters and RV amplitudes where drawn from the following distributions: 
\begin{description}
\item $P$: log-uniformly distributed over the interval $[10.0,365.25]$ d;
\item $K$: uniformly distributed over $[1.5,5.0]$ m s$^{-1}$;
\item $T_0$: uniformly distributed over the range: $[0,P]$;
\end{description}

Given the range of $K$ and the adopted value of $M_\star$, the corresponding interval of minimum planetary masses is between $\sim3$ $M_\oplus$ and 30 $M_\oplus$. 
All 10\,000 RV time series were generated with $N = 60$ observations uniformly distributed over $N_s = 3$. The season duration was set close to 6 months, 
with a daily observing window of approximately $12$ hr. 

\subsubsection{Single-planet eccentric orbits catalog}
\label{ce.cat}

The second catalog is composed of 10\,000 synthetic eccentric systems and their relative time series. 
The probability distribution function adopted for $e$ was the Beta distribution, following the recipe of \citet{kipping2013}:
\begin{equation}
\label{eq.pbeta}
\mathcal{P}_\beta (e;a,b) = {1 \over B(a,b)} e^{a - 1} (1-e)^{b-1},
\end{equation}
with $a = 0.867$ and $b = 3.03$. The remainder of the simulation setup was identical to that described in \S \ref{c.cat}.

\subsubsection{Multi-planet circular orbits catalog}
\label{d.cat}

\begin{figure}
\includegraphics[width=1.0\columnwidth]{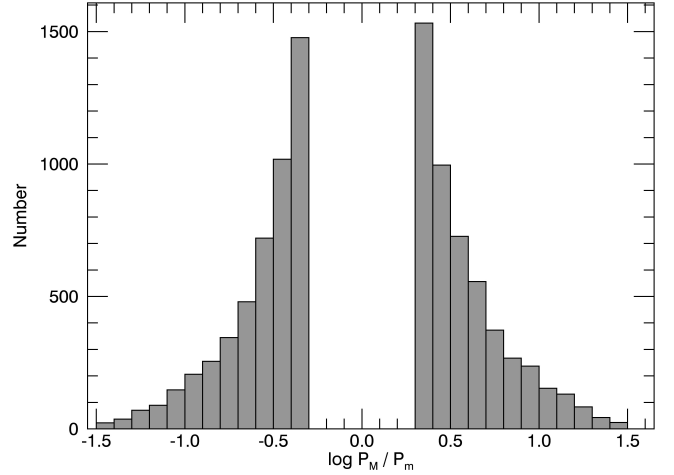}
\caption{\small Period ratio distribution function, with $P_\mathrm{M}$ and $P_\mathrm{m}$ the period of the planet with the larger and smaller amplitude, respectively.   \normalsize}
\label{tr.pc}
\end{figure}

The third catalog is composed of 10\,000 synthetic two-planet systems on circular orbits, and their relative time series. To generate each pair of companions, 
we first use the same $P$ distribution as in the first two catalogs, and then assign the orbital period $P'$ of the second planet following the distribution of 
period ratios observed for Kepler candidates by \citet{steffen2015}:

\begin{equation}
\mathcal{P}(\mathcal{R}) \propto \mathcal{R}^{-1.26},
\end{equation}
where $\mathcal{R} = P_\mathrm{o} / P_\mathrm{i}$, $P_\mathrm{i}$ and $P_\mathrm{o}$ being the periods of the inner and outer planet, respectively. 
The relation is valid for $\mathcal{R} \gtrsim 2$.
We do not require $P=P_\mathrm{i}$, so the probability density function for $P'$ is:
\begin{equation}
\label{eq.rapp}
\mathcal{P}(P';P)=
\begin{cases}
\left({P \over P'}\right)^{-1.26}, & \text{if $P' < P/2$,} \\
\left({P' \over P}\right)^{-1.26}, & \text{if $P' > P/2$.}
\end{cases}
\end{equation}
$P'$ was also required to be in the interval $[10.0,365.25]$ d. All other parameters in the simulated catalog were generated following the same prescriptions as 
in \S \ref{c.cat}. The resulting period ratio distribution is shown in Fig. \ref{tr.pc}.

\begin{figure}
\includegraphics[width=1.0\columnwidth]{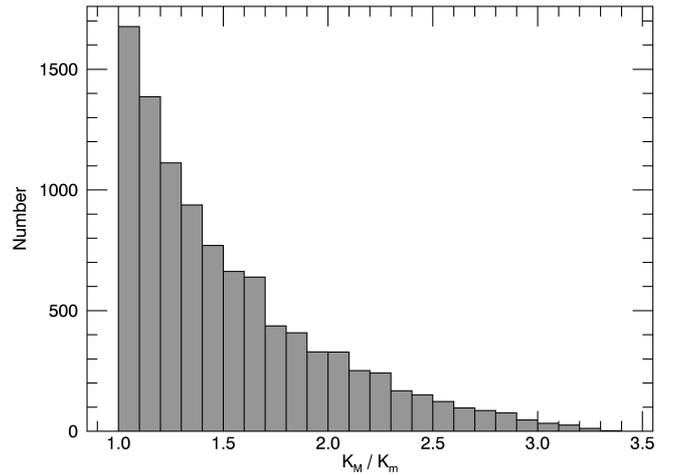}
\caption{\small Amplitude ratio distribution function, with $K_\mathrm{M}$ and $K_\mathrm{m}$ the larger and smaller amplitude, respectively.   \normalsize}
\label{tr.kc}
\end{figure}

We denote the largest and smallest amplitude $K_\mathrm{M}$ and $K_\mathrm{m}$ respectively, and the corresponding periods $P_\mathrm{M}$ and $P_\mathrm{m}$.
The distribution function of amplitude ratios is shown in Fig. \ref{tr.kc}.

\subsubsection{Multi-planet eccentric orbits catalog}
\label{de.cat}

The last catalog generated encompassed a set of 10\,000 eccentric two-planet systems, and their corresponding RV time series. As done in \S \ref{ce.cat}, 
the $e$ values for both orbits were drawn from the Beta distribution \citep{kipping2013}. In order to avoid unrealistic configurations corresponding to 
clearly dynamically unstable orbits, the masses, orbital separations, and eccentricities of a pair of synthetic planets were generated in order to fulfil 
the analytic Hill-stability criterion \citep[and references therein]{giuppone2013}:
\begin{equation}
\label{stab.crit}
 \left( \mu_1 + \mu_2 { a_1 \over a_2} \right) \left( \mu_1 \gamma_1 + \mu_2 \gamma_2 \sqrt{a_2 \over a_1} \right)^2 > \alpha^3 + 3^{4/3} \mu_1 \mu_2 \alpha^{5/3},
\end{equation}
with $\mu_i = m_i / m_\star$, $\alpha = \mu_1 + \mu_2$, $a_i$ the semi-major axis of planet $i$ and $\gamma_i = \sqrt{1-e_i^2}$. Systems violating this criterion were discarded.

Since the stability criterion penalizes highly eccentric orbits, in order to avoid a statistically insignificant sample of highly eccentric wide systems, we cut 
the eccentricities distribution at the $e =0.5$ level, which includes roughly $90\%$ of the systems. 

In order to study the sensitivity to the $P/2$ harmonics of eccentric orbits, we raised the period ratio lower limit in Equation \ref{eq.rapp} 
to $\mathcal{R} =2.5$, to avoid overlapping with signals from planets in 2:1 resonance.

\section{Results}
\label{sing.res}

The comparative study of the efficiency of the three period search algorithms presented here is carried out applying sequentially GLS, BGLS, and FREDEC to 
each of the four simulated datasets described in \S \ref{synt.cat}. Indeed, other studies in the past (e.g., \citealt{walker95}; \citealt{cumming99}, \citeyear{cumming08}; 
\citealt{endl02}, \citeyear{endl06}; \citealt{nelson98,eisner01,cumm04,narayan05,bonfils13,faria16}) 
have focused on gauging the sensitivity of RV planet searches to single-planet architectures utilizing periodogram analysis tools applied to synthetic as well 
as actual datasets in a variety of situations (large/small number of observations, periods shorter/longer than the duration of the observations, small and large companion masses). 
The systematic performance evaluation of GLS, BGLS, and FREDEC in the single planet case is useful in this context as it provides the opportunity to define and train on grounds 
that are better understood the comparison metrics to be used later for the comparative analysis of multiple circular and Keplerian signals, which has not been investigated in the past. 

For the purpose of maximizing the homogeneity of the analysis, we have set the maximum value of FAP considered for evaluation of a signal at 10\%, 
driven by the in-built FAP $ < 0.1$ limit in FREDEC \citep[see][section 4.2]{baluev2013fre}. For GLS, the FAP has been calculated following Eq. 24 and 25 in \citet{zechkur2009}. 
For BGLS, we followed \citet{mortier2015} and adopted as FAP value the relative probability between the two highest peaks. In practice, statistically significant 
detections are considered only those with FAP below the threshold FAP$_\mathrm{thr} = 1\times10^{-3}$.

To further quantify the quality of the results of the different algorithms we also calculated for each time series the true fractional error between the best output period 
$P_\mathrm{out}$ and the true simulated one $P_\mathrm{in}$: 
\begin{equation}
\label{eq.dp}
\Delta P = {P_\mathrm{in} - P_\mathrm{out} \over P_\mathrm{in}},
\end{equation}
and considered a correct identification of a given period when $\Delta P < 0.1$. For FREDEC we considered a planetary system as correctly identified if all the input periods were recovered in the output set with a fractional error lower than $10\%$, even in the presence of additional output periodicities, as well as we considered as wrong solutions that did not contain the input periods, even if they contained some of their harmonics.

To compare the algorithms we describe their performances by means of two global performance metrics: the completeness $C=N_\mathrm{corr}/N_\mathrm{cat}$ identifies the fraction of 
correctly identified planets signals $N_\mathrm{corr}$ with respect to the total simulated planets in the catalog $N_\mathrm{cat}$; the reliability $R=N_\mathrm{corr}/(N_\mathrm{corr} + N_\mathrm{FP})$ is 
the ratio of correct detections to the total of correct plus false alarms $N_\mathrm{FP}$. Finally, we quantify dependencies of the performance on the relevant parameters by using simple scaling relations expressing, for 
example, the detection efficiency as a function of the ratio $K/\sigma$ between planetary signal amplitude and single-measurement uncertainty. All the analysis is carried out 
using FAP $<$ FAP$_\mathrm{thr}$. 

\subsection{Sanity check on white noise}
\label{wn.res}

\begin{figure}
\includegraphics[width=1.0\columnwidth]{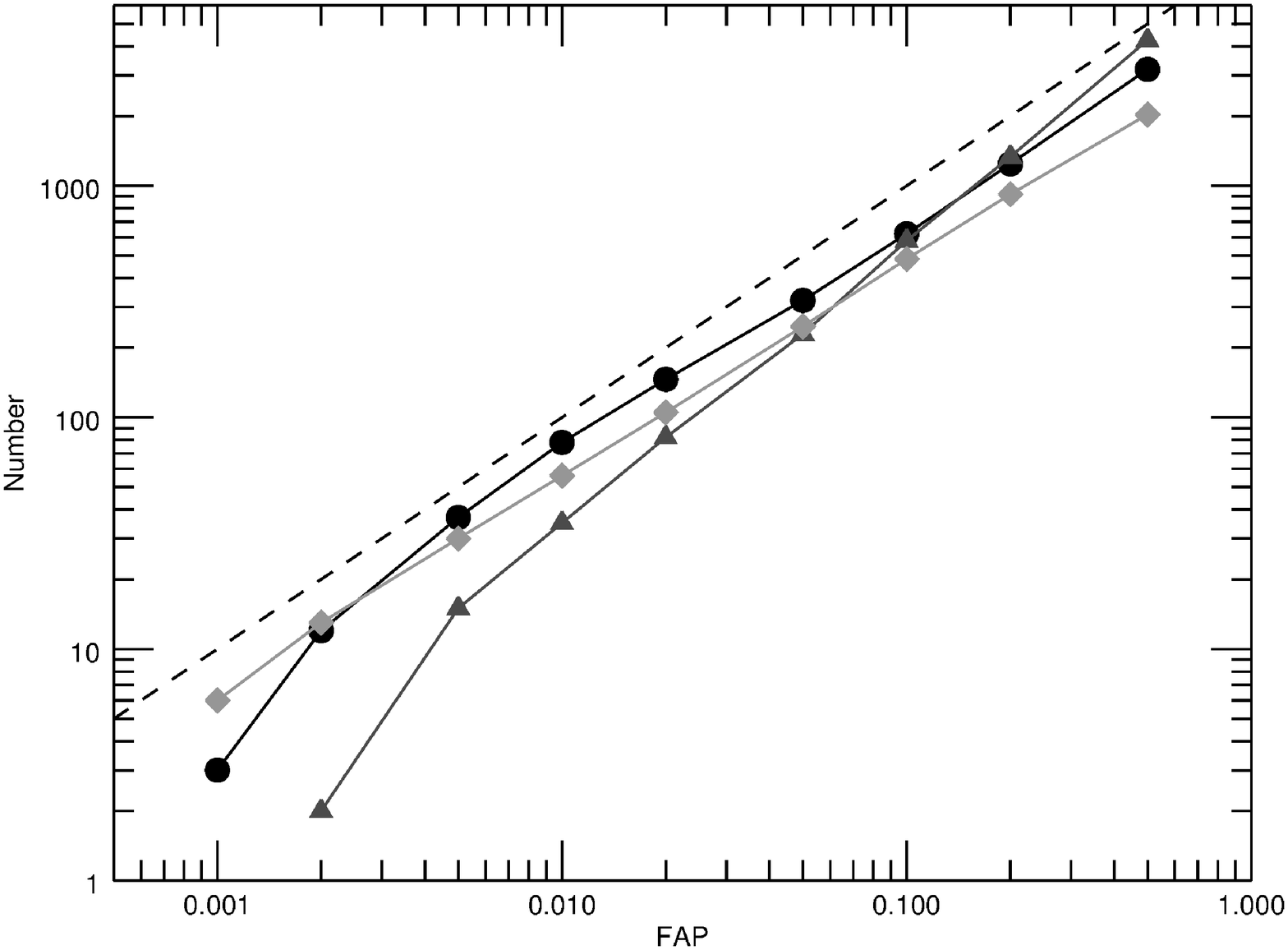}
\caption{\small Number of false positives found in 10\,000 white noise realizations as a function of the FAP threshold: 
black circles for the GLS, dark grey triangles for BGLS, and light grey squares for FREDEC. The dashed line is the theoretical expectation. \normalsize}
\label{wn.fap}
\end{figure}

The standard experiment to gauge the false alarm rate in the presence of pure white noise due to the statistical FAP threshold adopted for each 
algorithm should give expected results (e.g. 1\% of false positives for a FAP of 1\%). We have generated 10\,000 time series with pure white noise, $N=60$, and 
$N_s=3$, and run the three algorithms sequentially. We show in Figure \ref{wn.fap} the fraction of false alarms as function of FAP threshold.

We can see that all three curves are systematically lower than the dashed line, corresponding to the ideal relation between FAP and number of false positives. All 
three algorithms appear robust against false positives, within the limits of the FAP definition for each method.

\subsection{Single-planet circular orbits catalog}
\label{res.circ}

We applied GLS, BGLS, and FREDEC on the circular orbits catalog computing the periodograms at $10^3$ logarithmically spaced periods over the interval $[1,10^3]$ d.

\begin{table}
\caption{Circular orbits catalog results}
\label{tab.circ}
\begin{tabular}{lccc}
\hline
& $C$ & FP fraction & $R$ \\
\hline
GLS & $94.0 \%$ & $0.3 \%$ & $99.6 \%$ \\
BGLS & $87.9 \%$ & $0.0 \%$ & $100.0 \%$\\
FREDEC & $87.8 \%$ & $0.4 \%$ & $99.6 \%$\\
\hline
\end{tabular}
\end{table}

In Table \ref{tab.circ} are shown the overall $C$ and $R$ values for the three algorithms, along with the fraction of false positive signals found in the catalog. 
All methods show very high $C$ values, GLS performing slightly better ($\sim6\%$) than BGLS and FREDEC. Reliability levels are virtually at 100\% for all 
methods, given the extremely low fraction of false positive signals. 
There is however a significant discrepancy in the level of concordance between the three methods, that is the fraction of detected systems that is common: only 80\% of all detected signals is in common between GLS, BGLS, and FREDEC. 
These effects are best understood by looking at the structure of the dependence of the FAP on $K/\sigma$ in the three cases.  

\begin{figure*}
 \begin{center}
$\begin{array}{cc}
\includegraphics[width=1.0\columnwidth]{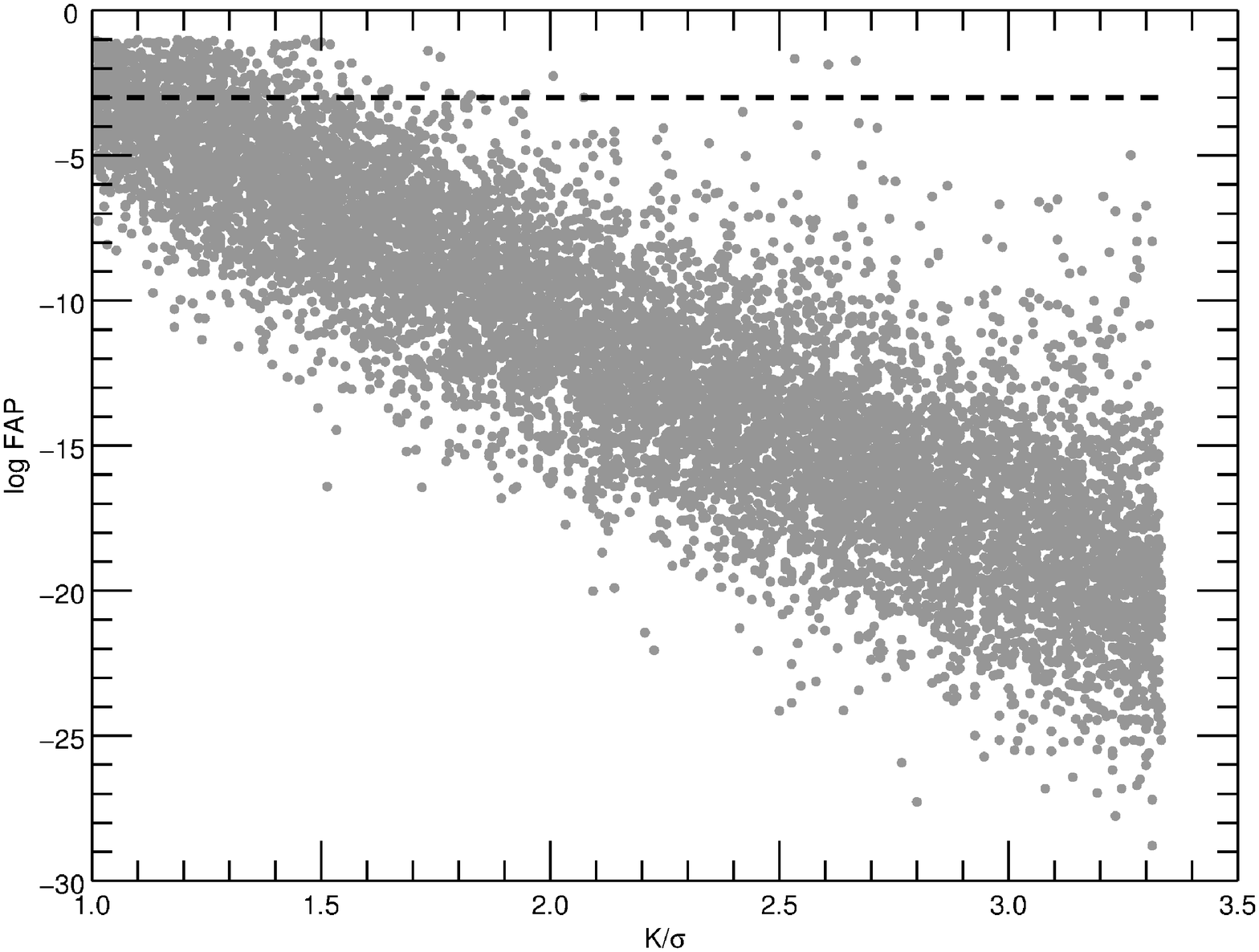} 
\includegraphics[width=1.0\columnwidth]{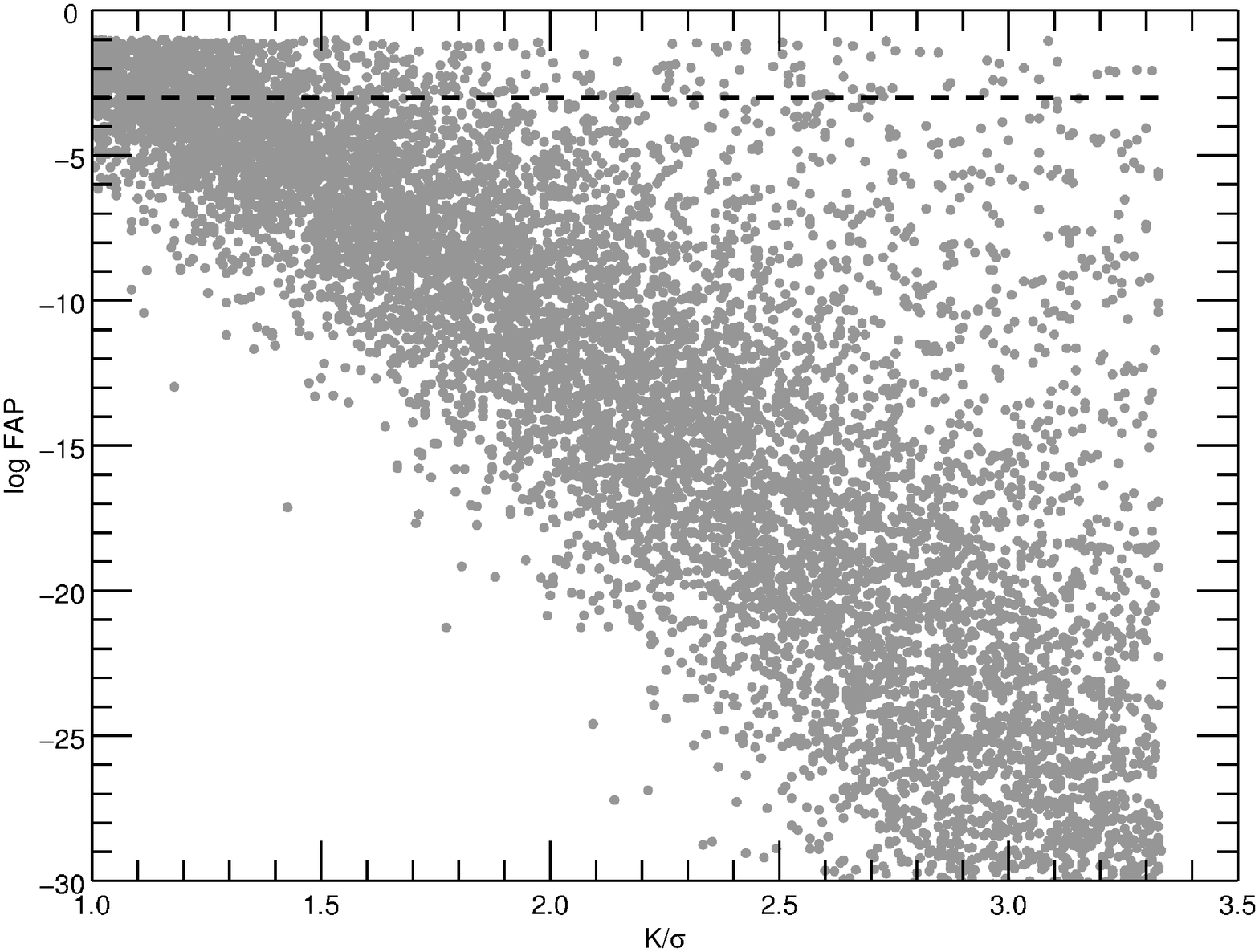} \\
\includegraphics[width=1.0\columnwidth]{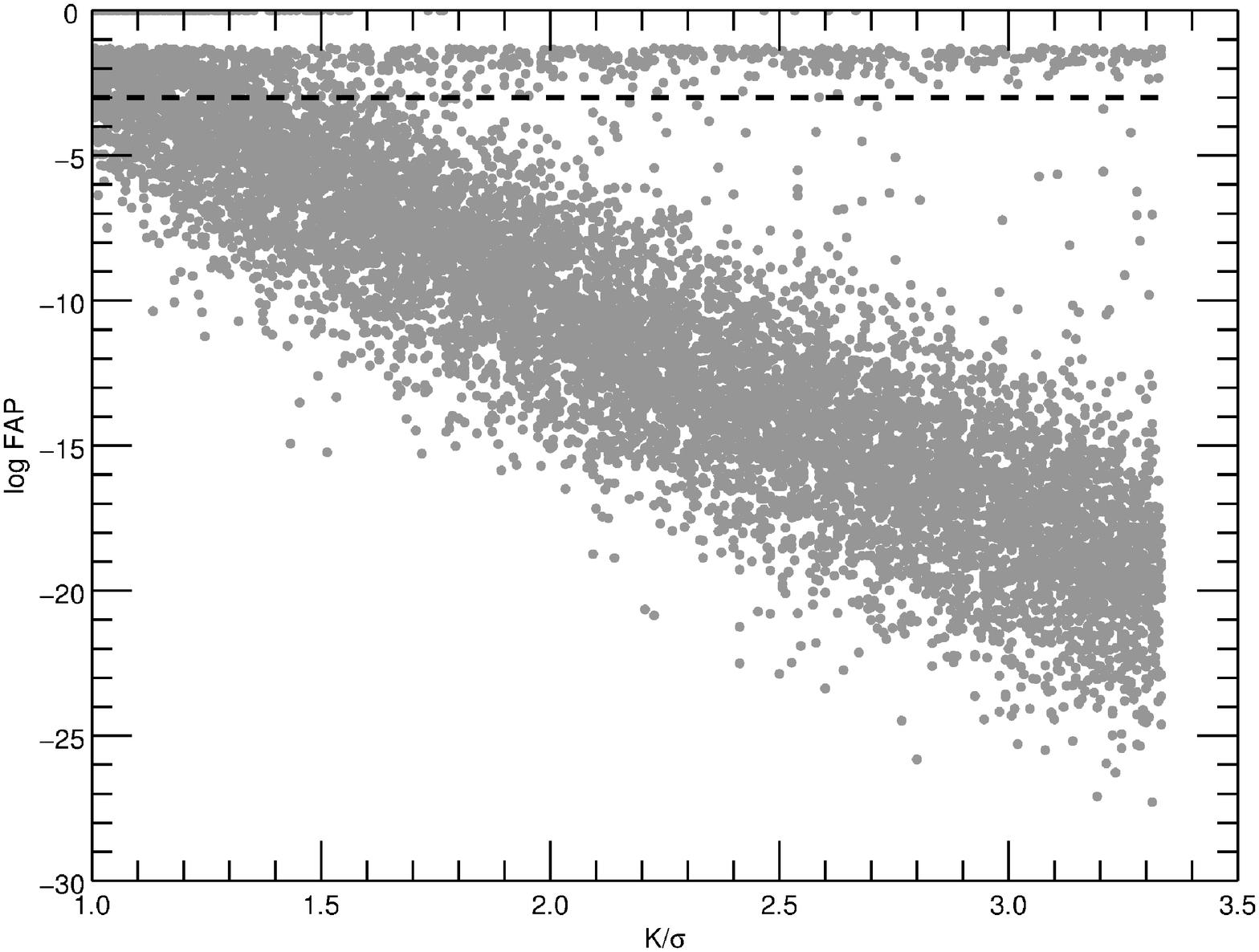} 
\includegraphics[width=1.0\columnwidth]{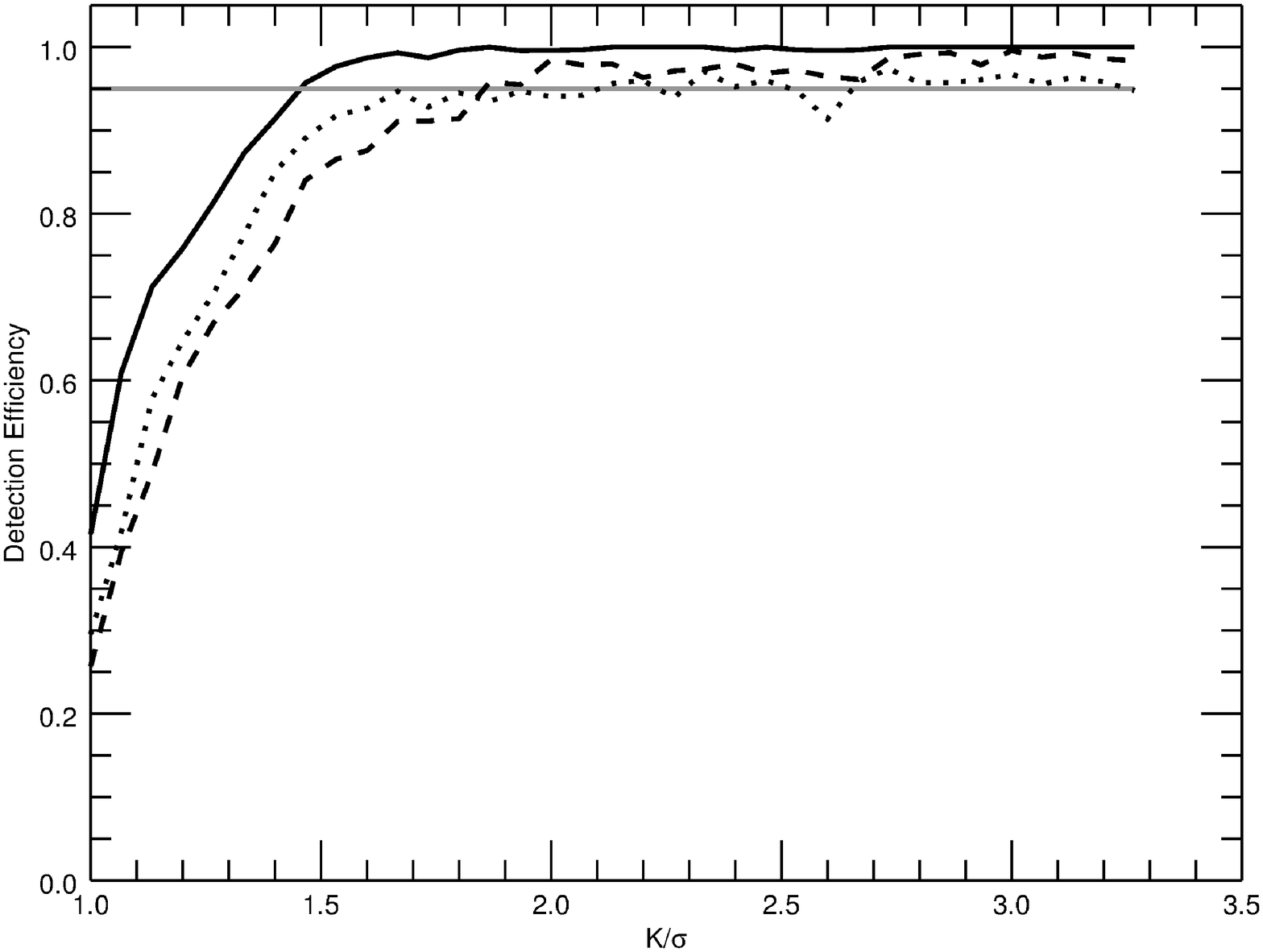} \\
\end{array}$    
\end{center}
 \caption{\small Dependence of the FAP on the $K/\sigma$, for top left) GLS, top right) BGLS and bottom left) FREDEC applied to the circular orbits catalog.  
 The black dashed line represents the $10^{-3}$ FAP level. 
 bottom right) Detection efficiency as a function of $K/\sigma$, for the circular orbits catalog. The solid back line is for GLS, the dashed black line for BGLS, 
 and the dotted black line for FREDEC. The grey solid line indicates the $95\%$ level of detections with FAP $<$ FAP$_\mathrm{thr}$. \normalsize}
\label{c.fa}
\end{figure*}

As shown in Fig. \ref{c.fa} (upper two panels and bottom left panel), the FAP decreases approximately log-linearly with increasing $K/\sigma$, as expected, 
BGLS highlighting a steeper dependence, and much larger spread in (statistically significant) FAP values in any given bin in $K/\sigma$. 
Furthermore, we notice that for BGLS very high FAP values are obtained even for $K/\sigma\gtrsim3$, which is not the case for GLS. 
FREDEC also highlights a systematically different behaviour with respect to GLS, stemming from its simultaneous multi-frequency identification approach. 
In this case, the small fraction of high-FAP systems that is recorded, independently of $K/\sigma$, corresponds to systems in which more than 1 signal is 
identified by FREDEC. No such cases are seen below the FAP$_\mathrm{thr}$ level. 

The bottom right panel of Fig. \ref{c.fa} quantifies the dependence of detection efficiency on $K/\sigma$. For GLS, $K/\sigma\simeq1.5$ is enough for 
correct recovery of the signals with $>95\%$ efficiency, while this result is achieved by BGLS at $K/\sigma\simeq2.0$. Unlike the other two methods, 
FREDEC never reaches close to the 100\% efficiency level, due to the systematic effect described above, that identifies $\approx5\%$ of low-FAP systems, 
independently of $K/\sigma$. Overall, GLS appears $\sim10\%$ more efficient than the other two algorithms, even in the limit of $K/\sigma\approx1$. 
The results obtained here are in agreement with the findings of \citet{cumm04}, but highlight slight differences between the three algorithms. 

\begin{figure}
\includegraphics[width=1.0\columnwidth]{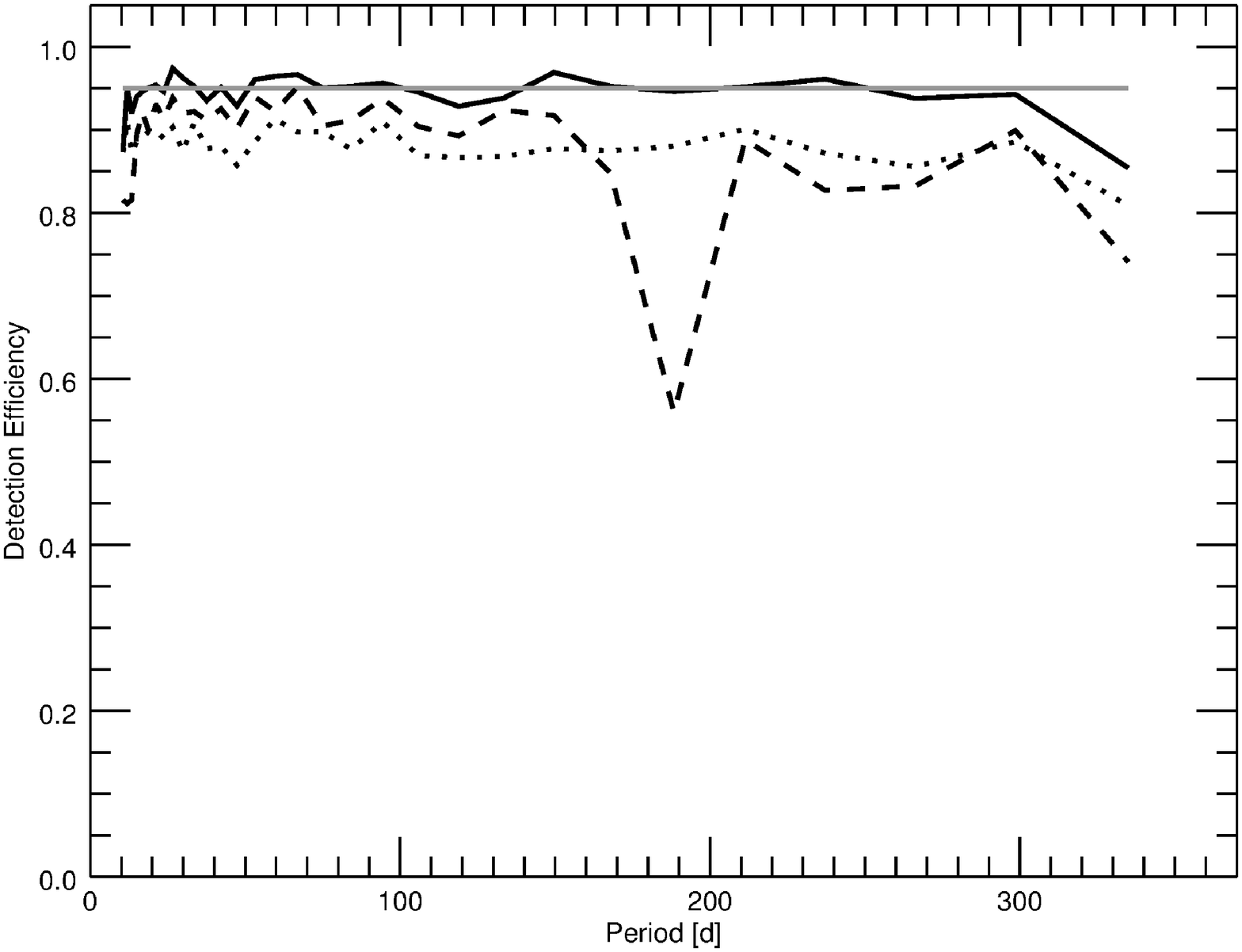} \\
\caption{\small Dependence of the FAP on the orbital period for the three algorithms applied to the circular orbits catalog. 
Line coding as in Fig. \ref{c.fa}\normalsize}
\label{cgb.fp}
\end{figure}

We show in Fig. \ref{cgb.fp} the behaviour of FAP with $P$ for GLS, BGLS, and FREDEC. No clear dependence of the FAP on the period of the detected 
signals is derived. This confirms the behaviour found by \citet{cumm04} using the LS periodogram coupled to a Keplerian fit, i.e. that the detection threshold is 
independent of $P$, for $P$ shorter than the time span of the observations. However, a clear loss in sensitivity for BGLS is seen for periods around $180$ d. 
This effect is related to the simulated length of the observing seasons, and is neither observed in GLS nor in FREDEC. 
The feature in correspondence of $\sim180$ d disappears from the BGLS analysis in the limit of higher sampling and unequal duration of each observing season (results not shown).

\subsection{Single-planet eccentric orbits catalog}
\label{ec.res}

All of the three algorithms fit pure sine waves\footnote{\citet{zechkur2009} presented also a fully Keplerian version of the GLS periodogram. The algorithm 
is signifantly heavier computationally than its circular version, and it would have required applying Keplerian fits to the data analysed with BGLS and FREDEC as well 
in order to keep homogeneity, thus making this study impractical given the available computational resources.}. 
We applied them (with the same boundaries in trial period as before) to a catalog of eccentric signals, 
to gauge their different biases and limitations (such as spurious detections of harmonics produced by eccentric signals) in the correct identification 
of $P$ and $K$ as a function of the eccentricity. In the analysis we distinguished between high and low eccentricity signals, the threshold being set to $e = 0.5$. 

\begin{table}
\caption{Eccentric orbits catalog results}
\label{tab.ecc}
\begin{tabular}{lccc}
\hline
& $C$ & FP fraction & $R$ \\
\hline
GLS & $86.1 \%$ & $0.9 \%$ & $99.0 \%$  \\
BGLS & $80.0 \%$ & $0.2 \%$ & $99.8 \%$ \\
FREDEC & $76.0 \%$ & $0.8 \%$ & $98.9 \%$ \\
\hline
\end{tabular}
\end{table}

Also in this case, we find that GLS and BGLS are in excellent agreement on the output values of the first periodogram analysis when both their signals are significant. 
Table \ref{tab.ecc} shows again $C$, $R$ and fraction of false positives of the different algorithms based on the analysis of the eccentric catalog. 
We can see that both $C$ and $R$ are lower than for the circular orbit catalog, while the fraction of false positives is higher. The behaviour of the individual 
algorithms is the same as before, with GLS being the most complete, and BGLS the most reliable. 
As expected, most of the incorrect identifications come from time series in which no significant period is found and/or those with particularly high eccentricity. We next take a 
closer look at the results of the individual algorithms. 

\begin{figure}
 \begin{center}
\includegraphics[width=1.0\columnwidth]{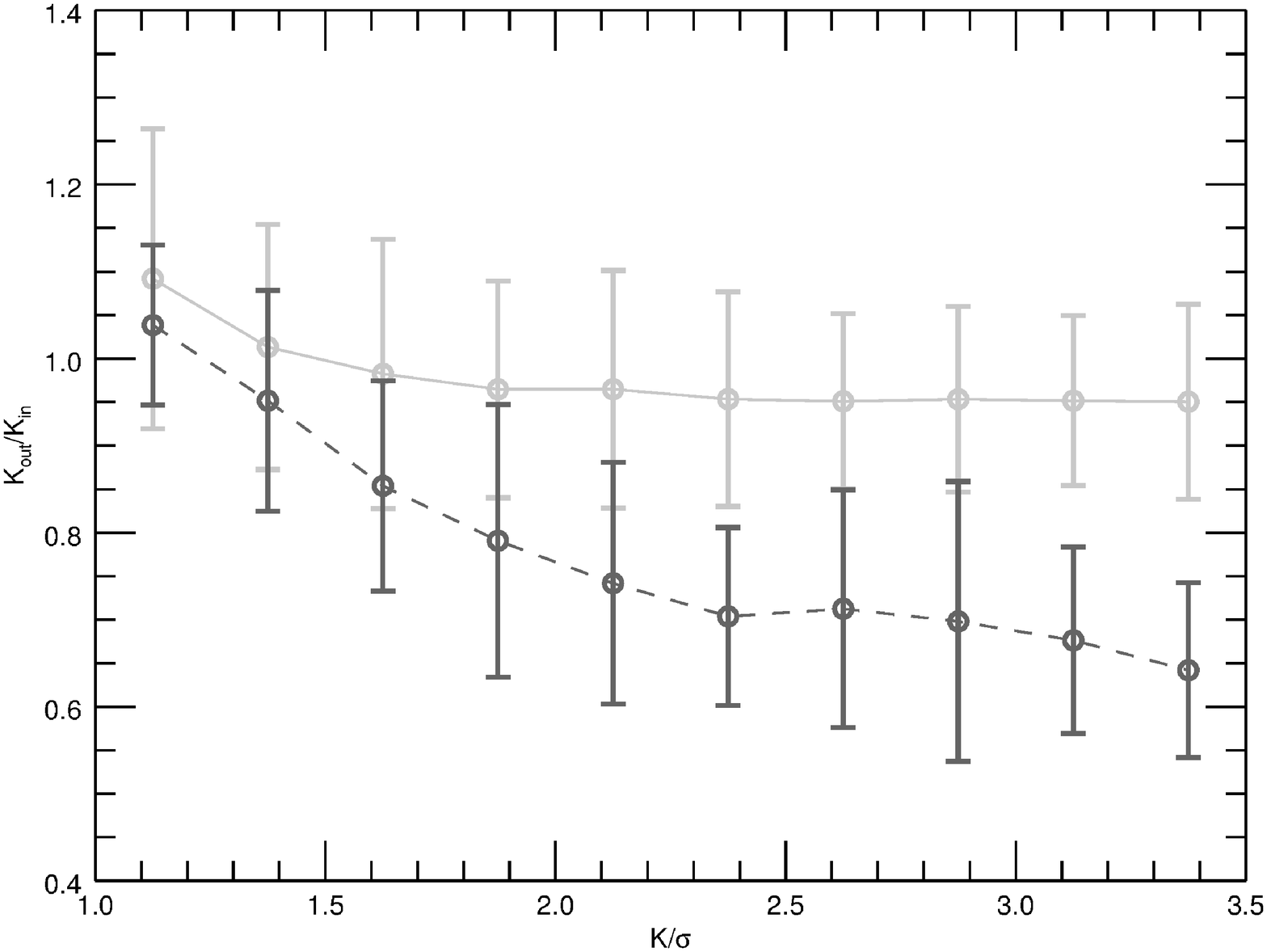} \\   
\end{center}
 \caption{\small The ratio $K_\mathrm{out}/K\mathrm{in}$ as function of $K/\sigma$, for the $e<0.5$ and $ e\geq0.5$ samples.  \normalsize}
\label{eg.piam}
\end{figure}

We show in Fig. \ref{eg.piam} the ratio $K_\mathrm{out}/K_\mathrm{in}$ of the fitted amplitude to the input K value expressed as a function of $K/\sigma$ 
for two regimes of eccentricity for GLS. The derived $K$ is systematically underestimated for the high-$e$ subsample. It is worth noticing that the result is opposite 
to that observed by \citet{shedwin08} in their analysis of eccentric RV signals. In that work, a systematic overestimate of the fitted $K$ values is a result of 
force-fitting Keplerian orbits with non-zero $e$ even in the limit of $K/\sigma\simeq1$, for which systematically large, and statistically not significant, eccentricities 
are obtained. The results for BGLS (not shown) are essentially identical. 

\begin{figure}
 \begin{center}
$\begin{array}{c}
    \includegraphics[width=0.95\columnwidth]{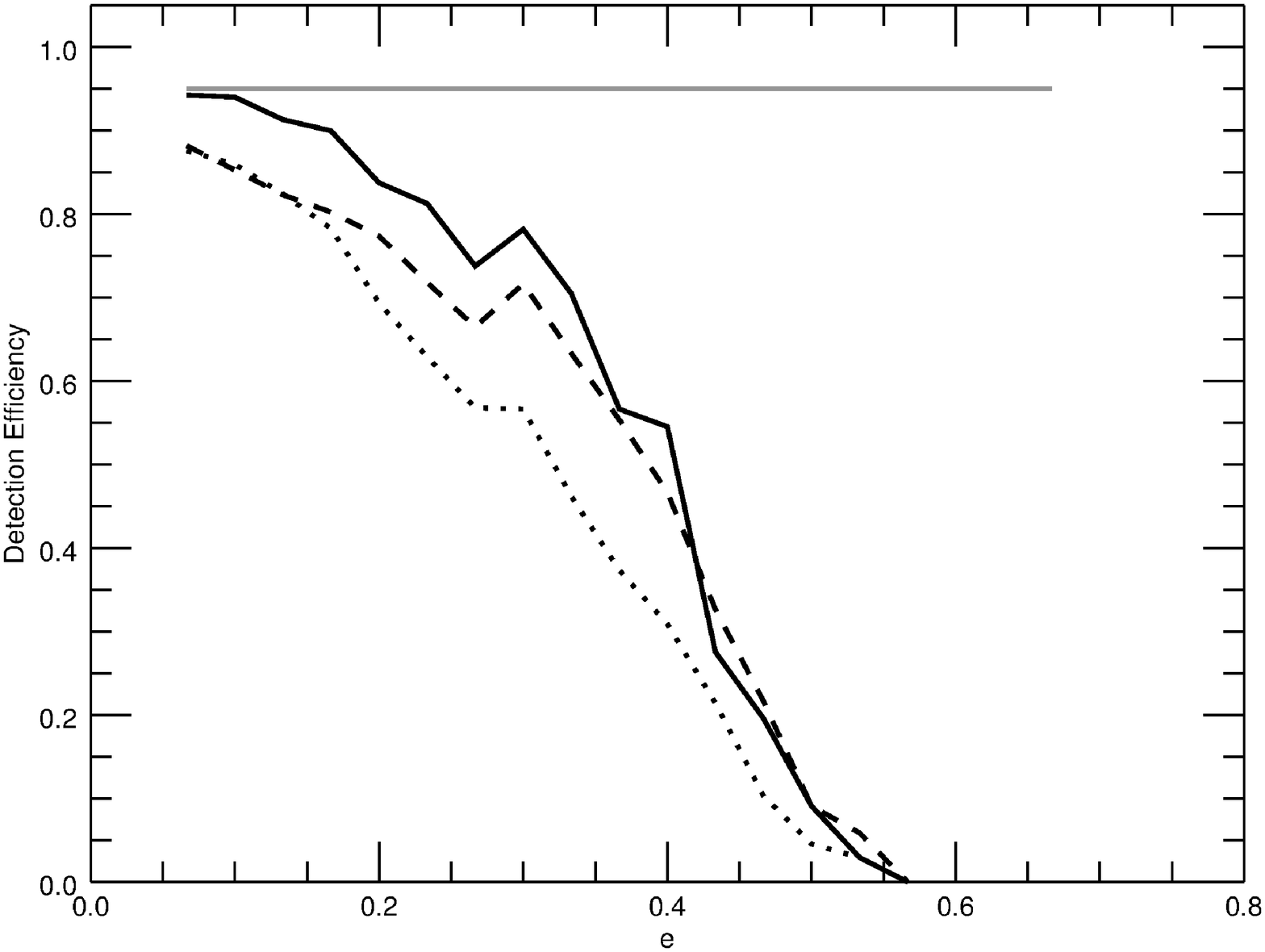} \\
    \includegraphics[width=0.95\columnwidth]{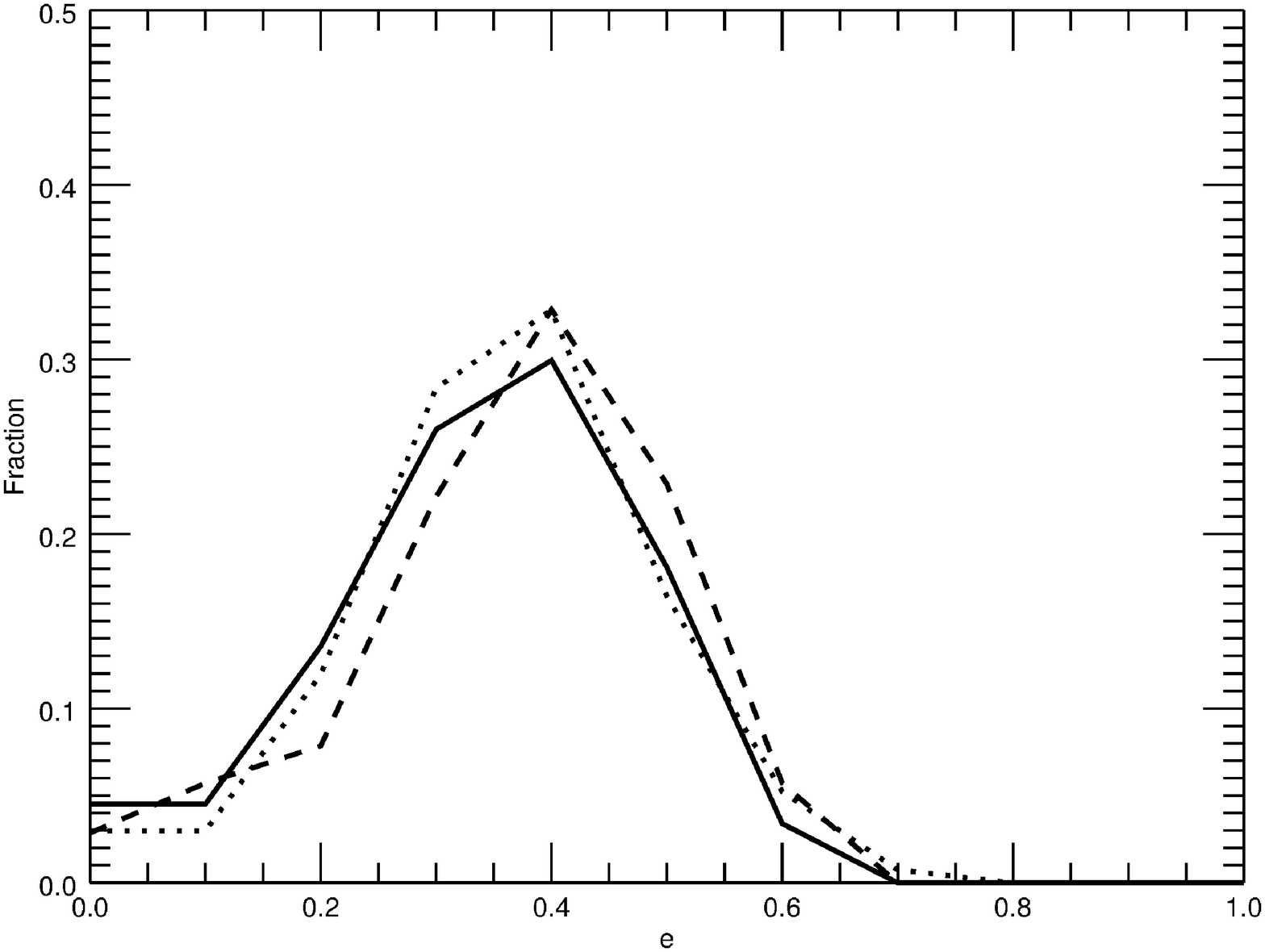} \\
\end{array}$    
\end{center}
\caption{Top: Detection efficiency above the $10^{-3}$ FAP threshold as function of eccentricity. The solid back line is for GLS, the dashed black line for BGLS, 
 and the dotted black line for FREDEC. The grey solid line indicates the $95\%$ level. Bottom: histogram of the 
fraction of significant periods identified in the residuals as function of eccentricity. Line coding is the same as in the upper panel.}
\label{ecceffplusres}
\end{figure}

\citet{cumm04} observed a quick decrease in detection efficiency for systems with $e \gtrsim 0.6$, finding that for too high eccentricities it is impossible to 
reconstruct the planetary signals. We derive in Fig. \ref{ecceffplusres} (top panel) a very similar result for all signal detection algorithms. For both GLS and BGLS 
detection efficiency drops to 50\% at $e\simeq0.4$, and no signals are detected (even with the largest $K/\sigma$ values) for $e\gtrsim0.6$. 
As for FREDEC, the behaviour is also similar to that of GLS and BGLS, with its detection efficiency reaching zero for $e\approx0.6$ (Fig. \ref{ecceffplusres}). 
However, an even steeper dependence of the algorithm on $e$ is seen, with the efficiency already lower by a factor of two with respect to GLS and BGLS at $e\simeq0.4$. 

As force-fitting a full Keplerian orbit to a low-amplitude signal often results in badly constrained (and artificially high) $e$ values, in practice signal subtraction is 
often carried out assuming a circular orbit. We carried out a GLS and BGLS analysis (with the same FAP thresholds as before) on the residuals to a 
circular-orbit fit to learn about the possible distortions in the time series induced by this approximation, particularly in the limit of high eccentricities for 
which residual power at first and higher order harmonics is expected. 

From the results of the residual analysis we notice that the fraction of significant signals found increases with increasing $e$ (Fig. \ref{ecceffplusres}, bottom 
panel), up to the eccentricity limit set by detection efficiency dropping to zero. For GLS, in $70\%$ of these systems the significant signal in the residuals is the 
first harmonic ($P/2$) of the input period. For BGLS this happens in $55\%$ of the cases. As for FREDEC, twice as many multiple significant signals are identified 
with respect to the circular orbit case. In this sample, the first harmonic at $P/2$ is found in $49\%$ of the cases, with a mean eccentricity of $\langle e \rangle = 0.41$ 
which is significantly higher than the average on the subsample and on the whole catalog. 

Finally, for all algorithms we tested whether increasing the length of the RV monitoring (up to 5 observing seasons) and/or doubling the number of observations per seasons 
(40 instead of 20) allowed to improve a) detection efficiency and/or b) mitigate the underestimation of the K value. No statistically significant changes in the behaviour 
shown in Fig. \ref{eg.piam} and Fig. \ref{ecceffplusres} were detected.

\subsection{Additional experiment: correlated noise}

As an additional experiment, we tested the performance of GLS and BGLS on a catalog with a more realistic stellar noise model. We added a simple correlated stellar activity signal, 
modeled with the analytical recipe by \citet{Aigrain12}. Our model considered 200 stellar spots, a realistic value for an M dwarf \citep{Barnes11}, and a rotation period 
of $30$ d; no differential rotation was included. We generated different spot distributions and sizes, in order to produce stellar activity signals with amplitudes $K_\star$
ranging between $1.5$ m s$^{-1}$ and 5 m s$^{-1}$. The planetary parameters were generated as in the circular orbits catalog of Sec. \ref{c.cat}.

We compared the results with an analogous catalog with the same planetary signals but no stellar activity, in order to quantify the decrease in detection efficiency of 
the planetary signals present in each time series. For both algorithms we used the same measure of relative detection efficiency 
utilized by \citet{vanderburg16} ($R_\mathrm{S/N}$, see their Eq. (1)). For GLS this is the square root of the ratio between the periodogram power measured with 
and without the stellar signal included, while for BGLS the quantity is the ratio between two Bayesian probabilities. 
An analogous experiment was not carried out using FREDEC, as no direct output in terms of periodogram power can be obtained from the software in its release. 

\begin{figure}
 \begin{center}
$\begin{array}{c}]
\includegraphics[width=1.0\columnwidth]{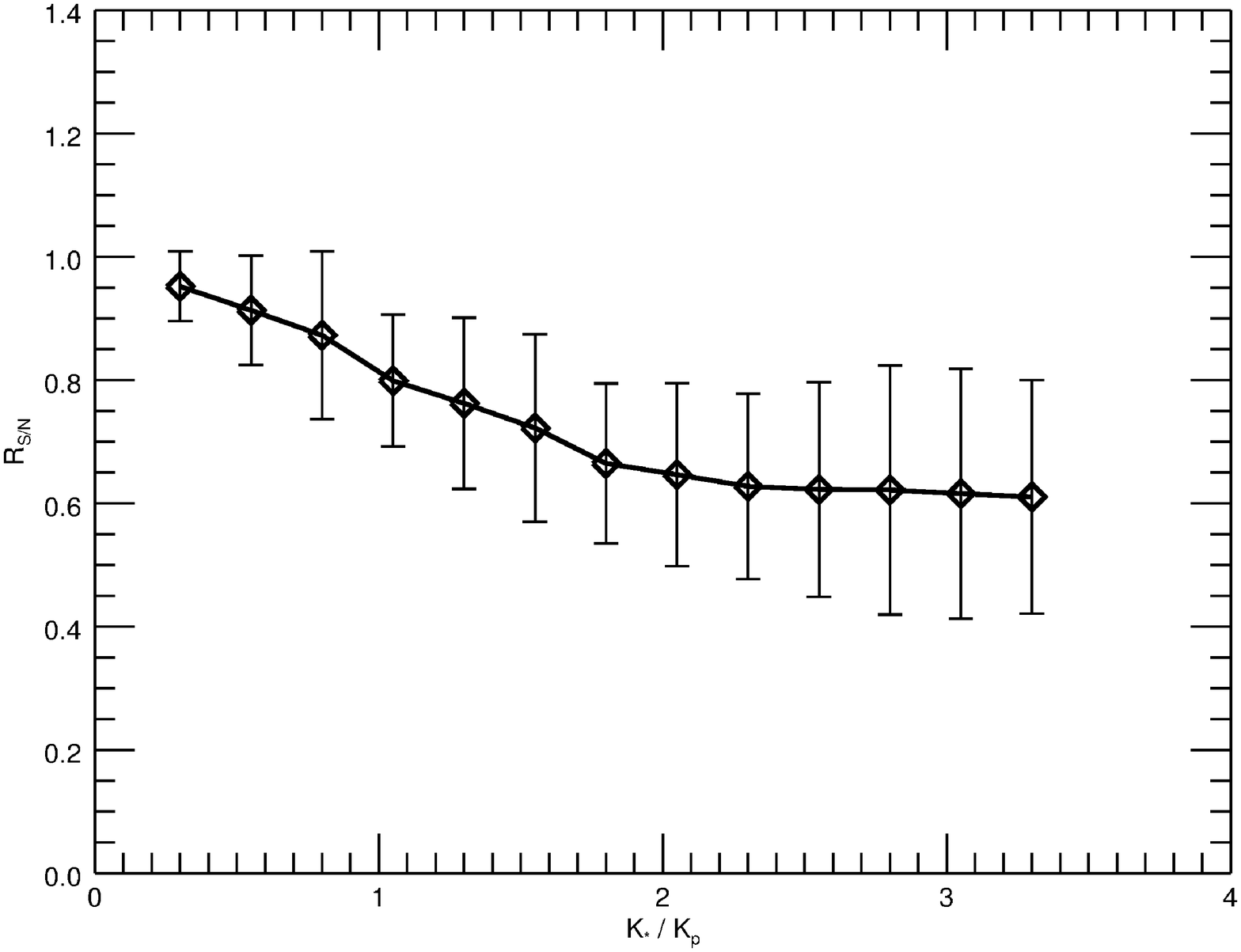} \\
\includegraphics[width=1.0\columnwidth]{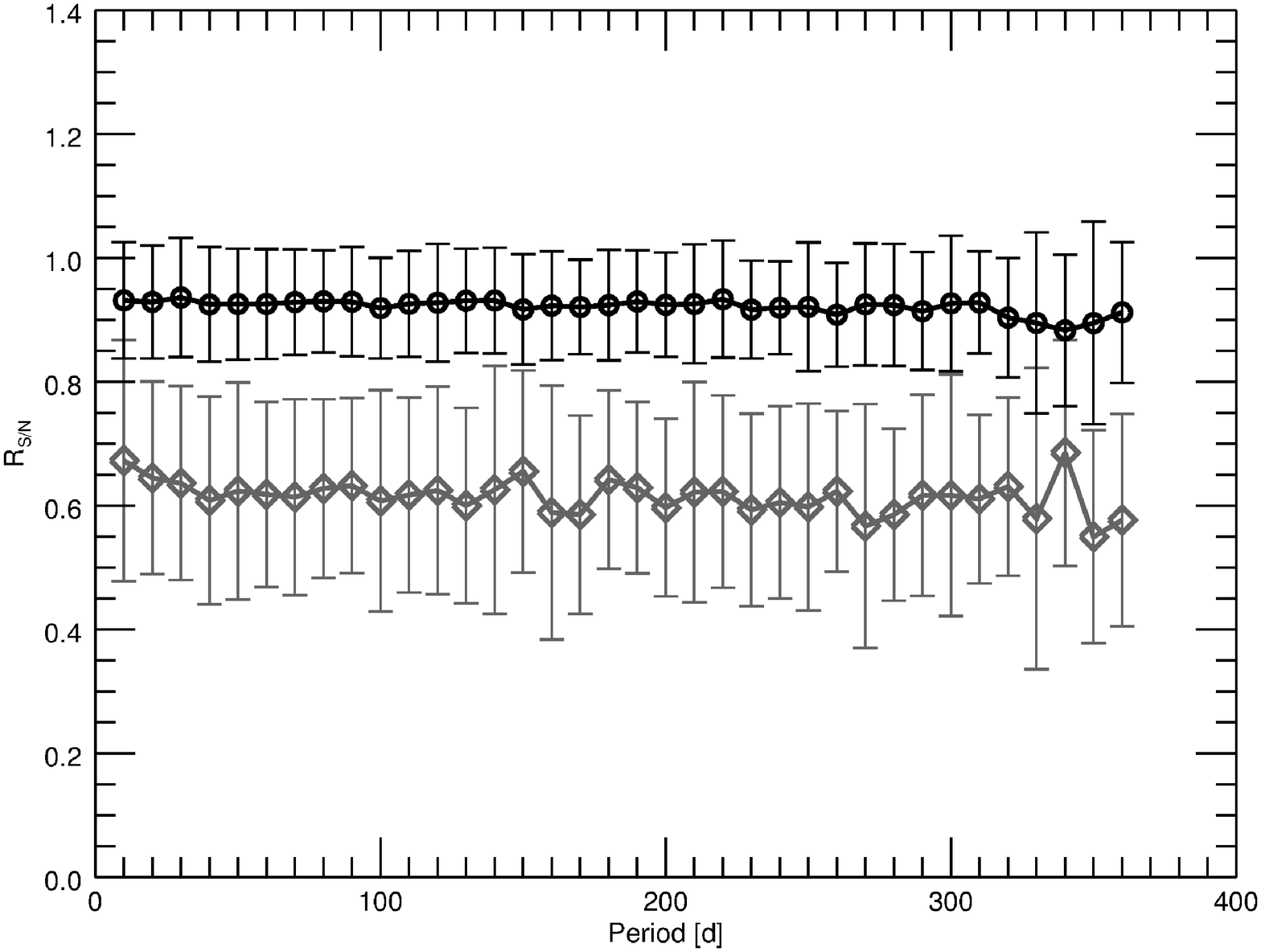}
\end{array}$    
\end{center}
 \caption{\small Top: Relative detection efficiency as function of the amplitude ratio $K_\star/K_P$. The dots and errorbars indicate the binned means and standard deviations. 
 Bottom: Relative detection efficiency as function of orbital period. The upper line is for the time series with $K_{\star}/K_P < 1$, the lower one for $K_\star/K_P > 2$. \normalsize}
\label{res.ehno}
\end{figure}

\citet{vanderburg16} found that the presence of correlated stellar noise produces a systematic degradation of $R_\mathrm{S/N}$ at all orbital periods investigated, 
with a stronger effect in the neighborhoud of the stellar rotation period and its first two harmonics. As we can see in the bottom panel of Fig. \ref{res.ehno}, 
our analysis using GLS confirms the systematic effect. Furthermore, the simulations allow us to quantify the dependence of the loss of detection efficiency as a function 
of the amplitude ratio $K_\star/K_P$ (kept constant at $K_\star/K_P=2$ by \citet{vanderburg16}). The result is shown in the upper panel of Fig. \ref{res.ehno}, 
in which we plot the relative detection efficiency as function of $K_\star/K_P$. We can see that for $K_\star/K_P\simeq 2$ the detection efficiency integrated over all 
periods drops by about $30\%$. 

The effect at the stellar rotation period and harmonics, discussed by \citet{vanderburg16}, is not present in our analysis. 
The drop in $R_{S/N}$ observed by \citet{vanderburg16} was due to the subraction of the fitted stellar activity signal 
from the RV dataset, translating in additional dilution of the planetary signal. Instead, we did not use any mock activity indicators to correct the RVs for the stellar signals, 
but simply studied the results of the periodogram analysis. It's also important to remember that the magnitude of the effect at $P_{rot}$ and its harmonics depends on the 
stellar spot configuration, but investigation of these aspects is beyond the scope of this experiment. 

The results with BGLS (not shown) follow similar trends, with the probability of the peak in presence of stellar activity being typically $10^2$ and $10^3$ times lower at 
$K_\star/K_P \simeq 2$ and $K_\star/K_P \simeq 3$, respectively.

\subsection{Multi-planet circular orbits catalog} 
\label{res.mult}

For GLS and BGLS, analysis of the multiple-planet simulations in the case of circular orbits proceeded (adopting the same periodogram setup as before) 
up to the period search in the RV residuals after removal of the second planetary signal. For FREDEC, up to 3 significant peaks were recorded. 

\begin{table}
\caption{Multi-planet circular orbits catalog results}
\label{tab.muci}
\begin{tabular}{lcccc}
\hline
& $C$ & FP fraction & $R$ \\
\hline
GLS &  73.1\% &  21.2\%  & 77.5\% \\
BGLS &  61.0\%  & 28.7\%  & 68.0\% \\
FREDEC & 72.8\%   & 8.5\%  & 89.5\% \\
\hline
\end{tabular}
\end{table}

We start by comparing directly the output periods of the GLS and BGLS algorithms. The first most significant period is identified by both GLS and BGLS 
in 100\% of the cases, thus both algorithms return the same results as in the single circular orbits catalog (see Section \ref{res.circ}). As expected, the same 
result is obtained in the analysis of the residuals after removal of the first and second significant periodicity, whenever the identified periods are the same 
for both algorithms (thus giving the same output structure of the post-fit residuals). 

As we can see in Table \ref{tab.muci}, the levels of complenetess and reliability for the correct detection of both injected planets are significantly lower than in the one 
planet case (see Table \ref{tab.circ}). Interestingly, BGLS shows the worst $C$ value for this catalog, thus proving its difficulties in dealing with multiple signals, 
as stated by \citet{mortier2015}. Both GLS and BGLS are prone to a large number of false positives, thus decreasing their $R$ value. While completeness for FREDEC is 
similar to that of GLS, its $R$ is significantly higher on the face of a much smaller number of false positives. This is likely due to the simultaneous multiple period 
search approach intrinsic to FREDEC. 

\begin{figure}
 \begin{center}
$\begin{array}{c}
    \includegraphics[width=0.95\columnwidth]{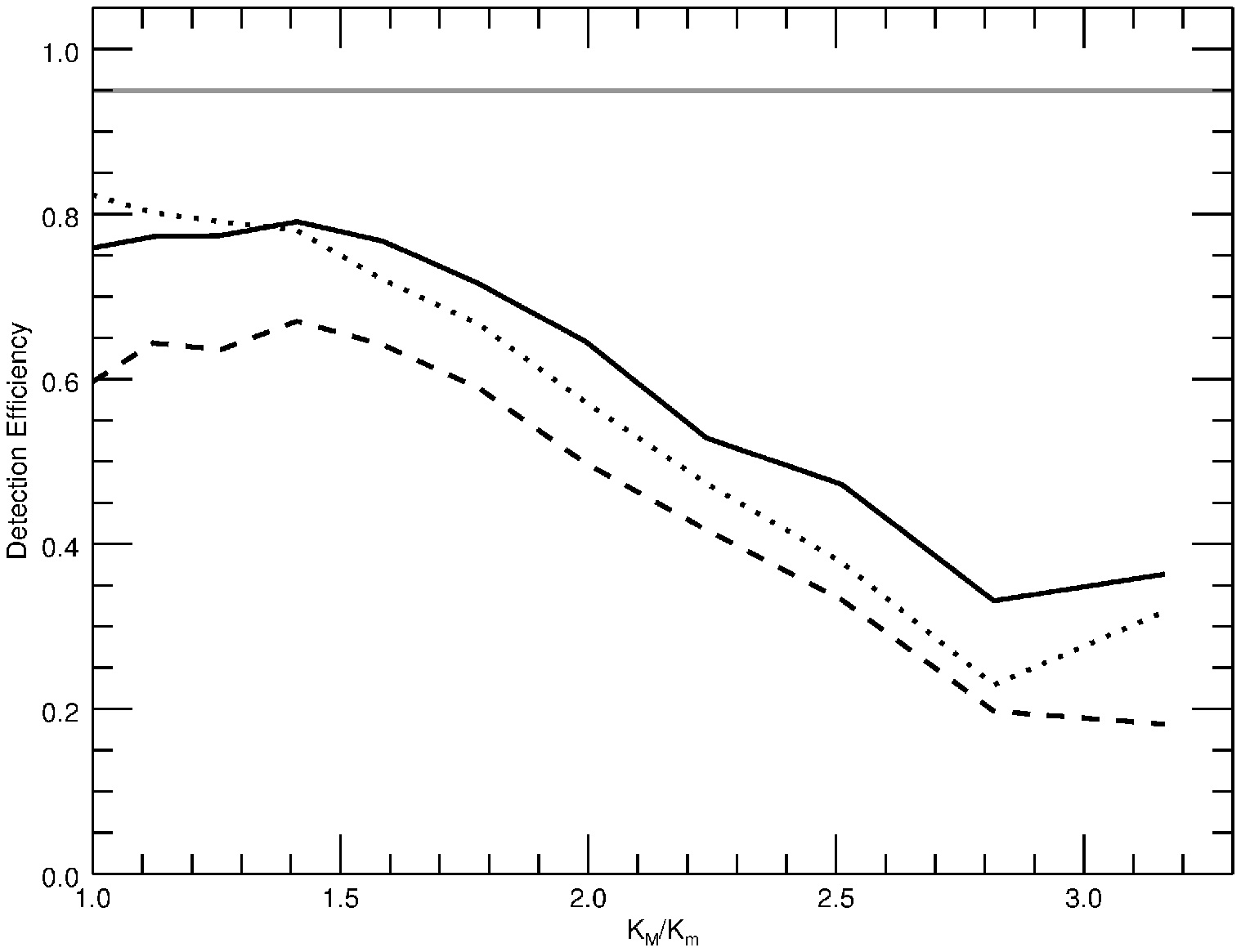} \\
    \includegraphics[width=0.95\columnwidth]{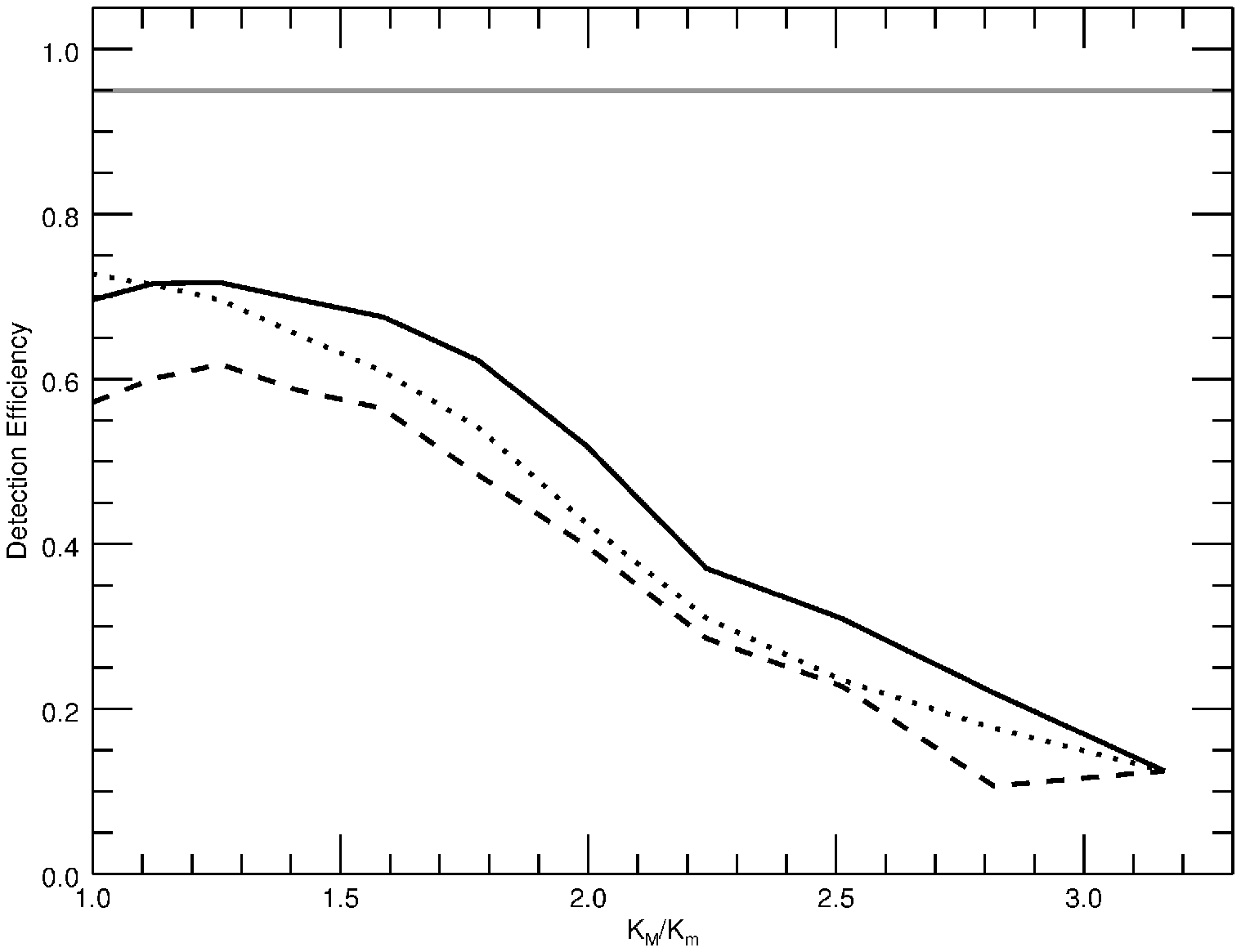} \\
\end{array}$    
\end{center}
\caption{Top: Detection efficiency above the $10^{-3}$ FAP in the multiple-planet, circular orbits case. The solid back line is for GLS, the dashed black line for BGLS, 
 and the dotted black line for FREDEC. The grey solid line indicates the $95\%$ level. Bottom: The same, for the multiple Keplerian orbits case. 
 Line coding is the same as in the upper panel.}
\label{deteffmulti}
\end{figure}

The top panel Fig. \ref{deteffmulti} captures, for three methods, the effect on the global efficiency of detection of both signals on the ratio of amplitudes 
$K_\mathrm{M}/K_\mathrm{m}$. Effiency never rises above $\sim80\%$ for either of the three algorithms. This value is maximum at $K_\mathrm{M}/K_\mathrm{m}\approx1$, the loss of $\sim20\%$ being 
due to the sample of systems with similar amplitudes, both close to the single-measurement precision. At $K_\mathrm{M}/K_\mathrm{m}\approx3$, efficiency is lower by a typical 
factor of 2 to 3, quantifying the difficulty in identifying correctly a second planet with $K_\mathrm{m}\simeq\sigma$ in the presence of a larger-amplitude signal, 
within the simulated observational scenario. Among the three methods, BGLS appears to suffer the most, performing typically a factor 1.3 to 2 with respect to GLS and FREDEC. 
We next turn to discuss some detailed features of the analysis carried out with each of the algorithms.

\begin{figure}
\includegraphics[width=1.0\columnwidth]{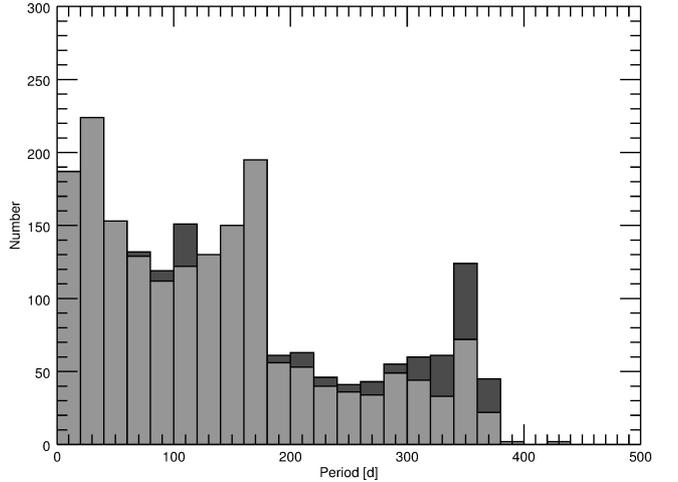}
\caption{\small Period distribution for GLS, on the multi-planet circular orbits catalog. The light grey area shows the output and the dark grey the 
strongest input signal, for the subsample with only one significant period identified.
\normalsize
}
\label{tg.piam}
\end{figure}

No significant signals are detected by GLS in $5.7\%$ of the systems. This occurs when both the input amplitudes are small, typically with $K/\sigma\lesssim1.8$ 
in both cases, and with the amplitude ratio being typically close to unity. There is no clear dependence on the periods, or their ratio. 
For 18.6\% of the systems only one significant period is identified. The input periods of this subsample 
are usually both long (typically $\sim 150$ d), and the ratio between the largest and the smallest amplitude is typically $\sim2$. In Fig. \ref{tg.piam} we show 
the period distribution for the output and input for this subsample: the distribution is almost the same, except for a clear aliasing effect for a significant 
fraction of systems with the strongest signal at $\sim 1$ yr, which are identified instead as being systems at 6 months of orbital period. 
The above results highlight some of the potential limitations for detection of these specific architectures of multiple-planet systems. 

GLS finds $2$ significant periodicities in $75.1\%$ of the time series. In the overwhelming majority of cases (96\%) two input signals are both identified correctly. 
In the remainder of the cases, incorrect identification of one or both periods is related to systems in which aliases created by the window function and its harmonics 
are detected. In only 0.65\% of the cases a third additional significant period is found after removal of the first two. This small sample is dominated by short-period aliases. 

In the BGLS analysis the fraction of 0, 1, 2, and 3 detected periods (with FAP $< 10^{-3}$) is 10.3\%, 27.5\%, 61.9\%, and 0.3\%, respectively. The global features of the sub-samples 
in the four cases are essentially identical to those discussed for the GLS cases. It is worth noticing the significant increase in null detections and in 
detections of only one period, which explains the lower $C$ value for BGLS in this experiment. 

In the FREDEC analysis the fraction of 0, 1, 2, and 3 detected periods (with FAP $< 10^{-3}$) is 18.7\%, 0.0\%, 80.8\%, and 0.6\%, respectively. The distributions of 
amplitudes and periods in the cases of no detections are similar to those of GLS and BGLS, although with somewhat larger average ratio of amplitudes and $K/\sigma\simeq1.5$ 
for the smallest of the two amplitudes in a system. The fraction of systems with two detected period is characterized by slightly longer periods and smaller amplitudes 
with respect to the GLS and BGLS cases, and a slightly lower fraction (90\%) of systems with both periods correctly identified is also recorded. 
Similarly to GLS and BGLS, incorrect identification of one or both periods is related to systems in which aliases are detected that are created by the window function and its harmonics. 
Contrary to GLS and BGLS, in the 0.6\% of cases with three significant periods detected, the sample is dominated by longer-period aliases (e.g., 1 yr). 

\subsection{Multi-planet eccentric orbits catalog}

\begin{table}
\caption{Multi-planet eccentric orbits catalog results}
\label{tab.muec}
\begin{tabular}{lcccc}
\hline
& $C$ & FP fraction & $R$ \\
\hline
GLS  & 65.0\% & 26.3\%  &71.2 \%  \\
BGLS  & 55.2\% & 33.1\% & 62.5 \%  \\
FREDEC & 62.2\% & 12.4\% & 83.4 \%  \\
\hline
\end{tabular}
\end{table}

The analysis of the multiple eccentric orbits catalog was performed as in the previous section. 
The complenetess and reliability levels of the algorithms are listed in Table \ref{tab.muec}, and as we can see are both lower than for the previous catalog. 
Again BGLS shows the lowest $C$ value, and also in this case FREDEC and GLS have comparable $C$ values but the former has higher $R$, and half has many false positives. 
The lower completeness levels translate in larger values of null detections and detections of only one significant period. 

The dependence of the number of detected systems on the main parameters (amplitude, period, eccentricity) follows generally the same 
behaviour observed in the previous experiments, for all methods. In particular, the mean amplitude increases with increasing number of signals 
found (as in Sec. \ref{res.mult} and the average eccentricity of both planets is lowest ($\sim0.15$) when all signals are correctly identified (as in Sec. \ref{ec.res}). 
As in the mutiple circular orbits case, the overall behaviour of detection efficiency for all three methods is mostly sensitive to the amplitude ratio, as demonstrated 
by the plot in the bottom panel of Fig. \ref{deteffmulti}. The impact of eccentric orbits is quantified in an additional efficiency loss of 10\%-20\%, slightly increasing 
towards larger $K_\mathrm{M}/K_\mathrm{m}$ values. 

There is however one difference: in the circular catalog the average period increased when more signals than in the input were found, while in 
this case it decreases. This is likely because the excess of signals recovered is due to poorly reconstructed orbits, which for the circular catalog is
 mainly due to not optimal sampling (and thus long periods), while for the eccentric catalog the extra signals found are also due to harmonics of the eccentric orbits, 
 whose impact becomes more significant with better orbit sampling. As a matter of fact, the fraction of cases when in addition to both the input signals one or both 
 the harmonics are detected is $72.2\%$, $77.8\%$, and $86.3\%$ for the GLS, BGLS, and FREDEC respectively, thus dominating over spurious detections.

\section{Summary and discussion}
\label{discuss}

In this paper we have carried out an extensive suite of numerical experiments aimed at a direct performance evaluation of three commonly adopted algorithms 
(GLS, BGLS, and FREDEC) in the search of significant periodicities in radial-velocity datasets, indicative of the presence of planetary companions. Using simple scaling 
relations (detection efficiency) and global performance metrics (completeness, reliability, false positives fraction) 
we have gauged the strengths and weaknesses of the three period search algorithms when run on a variety of classes of Doppler 
signals (one and two planets, circular and fully Keplerian orbits) of low amplitude ($1\lesssim K/\sigma\lesssim 3$), with representative 
realizations of observational strategies, different measurement noise prescriptions (simple Gaussian noise, stellar correlated noise), and adopting as reference an M dwarf 
primary. The main results can be summarized as follows: 

\begin{itemize}

\item The degree of completeness and reliability are very high for GLS, BGLS, and FREDEC in the single-planet, circular orbit case, with GLS being slightly more complete 
than the latter two methods. As a consequence, the fraction of false positives is very low. The overall detection efficiency is close to 100\% for all methods 
as long as $K/\sigma\gtrsim 2$, with a sharp decrease below $50\%$ in the limit $K/\sigma\simeq 1$. Also in this cases, GLS appears to be slightly ($10-15\%$) more efficient 
than BGLS and FREDEC in signal recovery when RV amplitudes get close to the single-measurement error;

\item The effect of eccentricity on correct signal identification by all methods is significant, as expected. A typical loss of 10\% in completeness is found, with GLS 
returning again the largest $C$ value. Reliability of detections remains however close 100\% given the mild increase in false detections. The latter are a clear function 
of increasing $e$, as long as detection efficiency remains above $\sim50\%$. The loss in efficiency of period recovery is a steep decreasing function of $e$, dropping 
to zero for all algorithms for $e\gtrsim0.6$. However, FREDEC shows a higher sensitivity to this parameter, with detection efficiency reduced by up to a factor of 2 in 
regime of intermediate $e$; 

\item A preliminary investigation of the levels of degradation of detection efficiency in presence of stellar correlated noise indicates effieciency losses of 20\% to 40\% 
in the range $1\lesssim K_\star/K_P\lesssim 3$ for GLS, and decrease of 2 to 3 orders of magnitude in the Bayesian probability of a detection for BGLS in the in the same 
$K_\star/K_P$ interval.

\item The difficulty in correctly identifying multiple planets is quantified through a typically reduced completeness level between 70\% (circular orbits) and 60\% 
(for Keplerian orbits), with BGLS performing slightly worse (10\%) with respect to the other two methods. Within the realm of the simulation scenario, and 
based on an analysis of the dependence of detection efficiency on the amplitude ratio $K_\mathrm{M}/K_\mathrm{m}$, the limitations 
induced by sub-optimal orbit sampling (particularly in the case of eccentric orbits) indicate as the most challenging architectures those containing 
signals with very similar amplitudes and $K\lesssim1.8$ m s$^{-1}$. In configurations containing two long-period companions with dissimilar amplitudes, the one with the lowest 
$K$ value is not detected in a significant fraction of cases (particularly for $K_\mathrm{M}/K_\mathrm{m}\gtrsim2$). Degradation in the degree of reliability is also clear, 
on the face of large fractions ($\sim30\%$) of false detections. In this respect, FREDEC appears more reliable than GLS and BGLS, with a false positive rate $\sim10\%$. 

\end{itemize}

The results presented in this paper complement and extend the comparative analysis of period search tools for planet detection in RV datasets carried out by \citet{mortier2015}. 
Our study encompasses a wide range of single-planet architectures, it includes a preliminary assessment of the effects of increasing levels of stellar correlated noise, 
and it addresses for the first time some of the complications induced by multiple-planet architectures. The most most important lessons learned are the following: 
1) even under idealized, best-case conditions (one planet, circular orbits, white noise, well-sampled orbits) different period search algorithms do not 
perform in an exactly identical fashion, 
particularly when it comes to the regime of signal amplitudes close to the single-measurement error; 2) in the presence of more complex signals, the most conspicuous 
element to underline is the different behaviour in the identification of false alarms: the standard approach of successive signals removal and investigation 
of the residuals (using GLS and BGLS) 
appears to be prone to as much as 3 times the amount of false positives obtained by an approach in which all statistically significant signals are searched simultaneously 
(using FREDEC), even in the idealized case of perfectly circular orbits. 

The analysis presented here is by no means exhaustive. Within the scope of this work, our results nevertheless underscore the 
urgent need for strenghtening and further developing sophisticated analysis techniques for the simultaneous identification of low-amplitude 
planetary signals in the presence of stellar activity. This is a crucial topic in the case of low-mass M-type hosts, for which stellar noise is often 
coupled to complex planetary RV signals induced by small-mass multiple systems, as testified by the significant literature presenting disputes on the nature, interpretation, 
and sometimes existence of multiple planets around some of our nearest low-mass neighbours (e.g., GJ 581, Kapteyn's star. See \citet{anglada16a}, and references therein). 
This is a particularly sensitive issue as M dwarf primaries constitute the fast track to the identification of potentially habitable terrestrial-type planets, whose 
abundance, albeit with large uncertainties, appears to be very high (e.g., \citealt{dreschar2013, kopparapu13}, \citealt{bonfils13}, \citeyear{bonfils2013}, 
\citealt{tuomi14, anglada16b})

It will be certainly necessary to use the largest possible set of observational constraints, including 
simultaneous photometric measurements for determining rotation periods and activity signals, and spectroscopic indicators and/or RV measurements at different wavelengths 
for mitigating and (hopefully) removing activity signals (e.g., \citet{vanderburg16}, and references therein). It will be equally important, however, to pursue 
aggressively advances in the path to the determination of the complete information content of RV datasets, via techniques that not only shy away from the standard 
residual analysis and implement global model fitting approaches (e.g., \citealt{hara16,dumusque2017}), but also through the application of improved methodologies for the simultaneous, 
robust identification of credible signals in time series (with very small fractions of false alarms), of which algorithms such as FREDEC constitute possible seeds. 
This necessity is expected to become pressing very soon, with facilities for ultra-high precision RV work such as ESPRESSO, that will seek to find (multiple) planetary signals 
with amplitudes even orders of magnitude smaller than other sources (primarily stellar in nature) of correlated RV variations.

\section*{Acknowledgements}

M.P. acknowledges the financial support of the 2014 PhD fellowship programme of INAF. 
M. Damasso acknowledges funding from INAF through the Progetti Premiali funding scheme of the Italian Ministry of Education, University, and Research.
We also thank the anonymous referee for the swift and useful review. 




\bibliographystyle{mnras}
\bibliography{biblio} 




\appendix

%


\bsp	
\label{lastpage}
\end{document}